# Advances in electrical contacts to single crystals of emerging materials for transport measurements

Huandong Chen[1,2*], Abhay N. Pasupathy[2,3], Jayakanth Ravichandran[1,4,5*]

[1]Mork Family Department of Chemical Engineering and Materials Science, University of Southern California, Los Angeles, CA, USA

[2]Condensed Matter Physics and Materials Science Department, Brookhaven National Laboratory, Upton, NY, USA

[3]Department of Physics, Columbia University, New York, NY, USA

[4]Ming Hsieh Department of Electrical and Computer Engineering, University of Southern California, Los Angeles, CA, USA

[5]Core Center of Excellence in Nano Imaging, University of Southern California, Los Angeles, CA, USA

*Email: hchen3@bnl.gov, j.ravichandran@usc.edu

## Abstract

Transport measurements that probe electrical resistivity of a material under varying external stimuli are among the most important experimental techniques in condensed matter physics. These measurements provide critical insights into the fundamental electronic properties of materials. In recent years, they have facilitated the discovery and exploration of intriguing physical phenomena and unique device functionalities in various emerging materials, particularly in the form of single crystals. However, unlike large-scale wafers or thin films, newly synthesized single crystals often pose substantial challenges in establishing reliable electrical contacts due to their irregular geometries, limited dimensions, inherent structural characteristics, and potential susceptibility to degradation. In this review, we highlight recent technological advancements in the fabrication of high-quality, lithographically defined multi-terminal electrodes on both exfoliable and non-exfoliable single crystals for transport measurements. Our work provides a practical guide for researchers seeking to select appropriate contact-fabrication strategies tailored to unique characteristics of emerging crystals.

## 1. Introduction:

Electrical transport measurements, enabled by the successful establishment of high-quality electrical contacts, represent one of the most extensively employed experimental techniques for discovering novel physical phenomena and demonstrating unique electronic functionalities in emerging materials.[1-3] In particular, single-crystal forms of materials, which are free of grain boundary-related defects, provide an ideal platform for probing intrinsic electrical properties closely tied to their crystallographic structures.[4-6] Additional advantages of transport studies on single crystals include the ability to investigate electronic anisotropy,[7-9] establish benchmark performance metrics,[10-12] and facilitate direct comparison with theoretical models.[13-15] Figure 1a shows several examples of recently developed single-crystal materials that have attracted significant research interest, including the flat-band lattice metal $Pd_5AlI_2$,[16] the heavy-fermion metal CeSiI,[17,18] the kagome metal $CsV_3Sb_5$,[19-21] the photoactive semiconductor $BaZrS_3$,[22,23] the phase-change chalcogenide $BaTiS_3$,[6,24,25] and the Dirac/Weyl semimetals such as $ZrTe_5$,[26-28] TaAs,[29-31] and $Cd_3As_2$.[32,33] Notably, these single crystals, typically synthesized through laboratory-scale techniques such as chemical vapor transport (CVT)[34,35] or molten-flux growth methods,[36,37] often exhibit irregular shapes and limited lateral dimensions ranging from a few hundred micrometers to several millimeters.

On the other hand, specific multi-terminal contact geometries are often essential for transport measurements under various external perturbations such as temperature, gate voltage, and magnetic field. Figure 1b and 1c depict several archetypal transport experiments alongside the commonly adopted contact geometries of devices. For instance, two-terminal configurations are typically employed for estimating resistivity when four-probe contacts are not feasible,[38] or for constructing rudimentary resistive switching devices such as photoconductors,[23,39,40] ferroelectric capacitors,[41,42] and memristors.[43,44] Three-terminal transistor geometries, implemented through either top or back gating, are commonly used to modulate the channel carrier densities.[45-47] Linear four-probe geometries are preferred for temperature-dependent resistivity measurements, particularly in studies of superconductivity[19,48] and phase transitions;[6,49] when configured in pairs, they can also provide insight into in-plane electronic anisotropy.[50,51] Additionally, the transmission line method (TLM) configuration serves to quantitatively assess contact resistance[52-54] and can be adapted to accommodate multiple two-terminal devices with varying channel lengths.[55] For accurate magnetotransport measurements, geometries such as van der Pauw[56-58] and Hall bars[59-61] are still crucial.

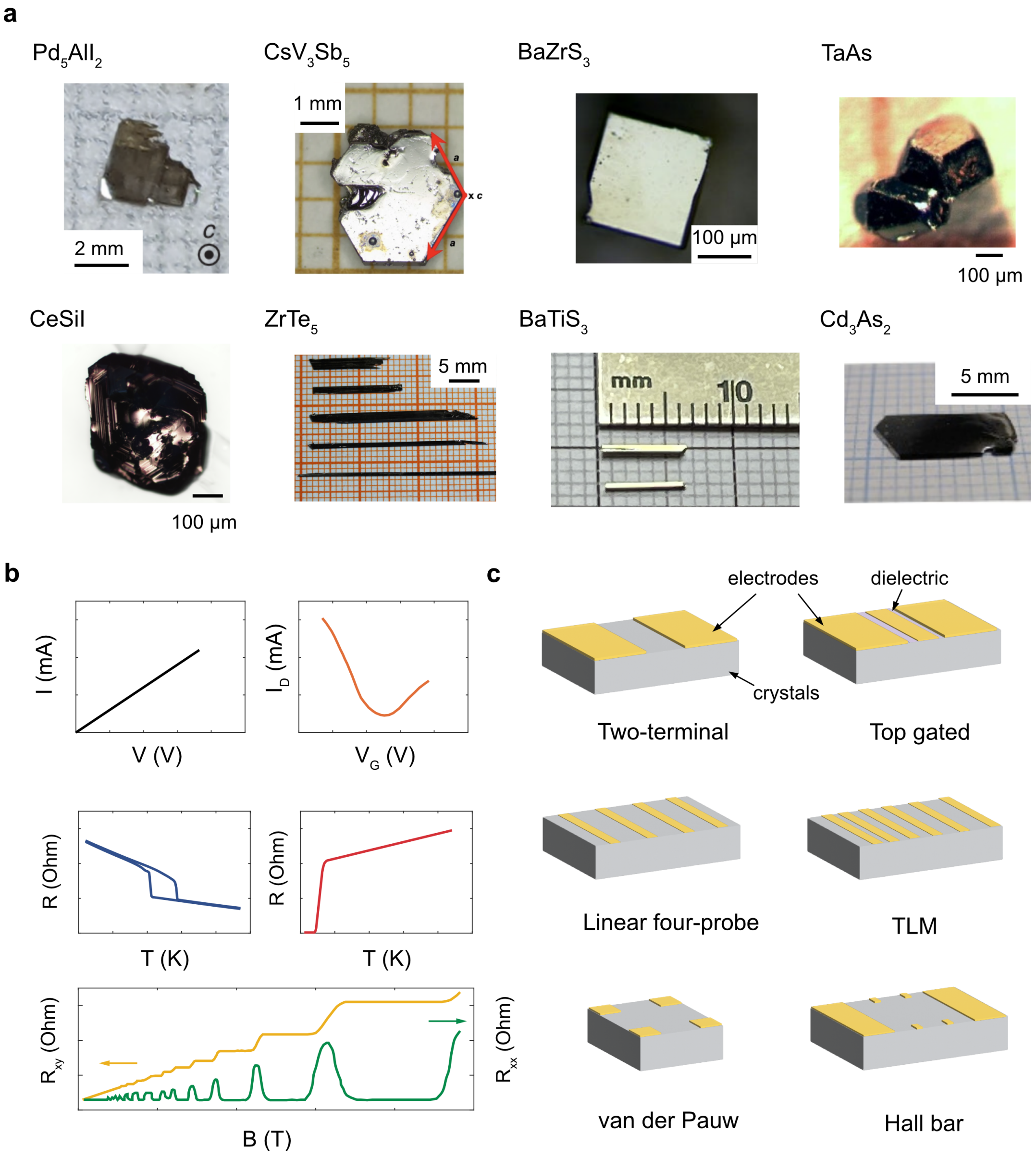

**Figure 1 Emerging crystals and transport measurements.** (**a**) Optical images of representative emerging single crystals. $Pd_5AlI_2$, Ref. [16] Copyright 2025, Springer Nature. CeSiI, Ref. [18] Copyright 2024, Springer Nature. $CsV_3Sb_5$, Ref. [21] Copyright 2022, Springer Nature. $ZrTe_5$, Ref. [28] Copyright 2017, Elsevier. $BaZrS_3$, Ref. [22] Copyright 2019, Springer Nature. $BaTiS_3$, Ref. [25] Copyright 2024, Springer Nature. TaAs, Ref. [31] Copyright 2015, Springer Nature. $Cd_3As_2$, Ref. [33] Copyright 2014, Springer Nature. (**b**) Typical electrical transport measurements of emerging crystals under various external perturbations such as temperature, gate voltage, and magnetic field. (**c**) Schematic illustration of representative device contact geometries adopted for various transport measurements.

In contrast to wafer-scale substrates or thin films, the fabrication of such versatile multi-terminal electrodes with precisely defined contact geometries on laboratory-grown single crystals is often nontrivial. Traditionally, researchers have relied on manually placing thin metal wires onto as-grown single crystals under an optical microscope and securing them with conductive adhesives, a method that remains one of the few viable solutions for single-crystal device fabrication.[3,62] While this technique has played a pivotal role in the discovery of superconductivity and phase transitions in various quantum materials since the 1970s,[63-65] it presents significant limitations. For instance, the method requires crystals of at least millimeter-scale dimensions and demands considerable user skill to manually position metal wires without shorting them to one another, thereby excluding many interesting but size-limited crystals from systematic transport investigations. Furthermore, the lack of precise control over contact geometries and of systematic contact optimization significantly limits the device quality. These issues are particularly pronounced in gapped systems such as semiconductors, where poor contacts can compromise device performance and, in some cases, lead to misinterpretation of transport behaviors by failing to resolve subtle variations in channel resistance.

Recent developments in single-crystal device microfabrication, particularly those enabled by lithography-based patterning and etching processes, have significantly advanced electrical transport studies of emerging single crystals.[6,30,66,67] In this review, we present a comprehensive overview of recent technological progress in fabricating high-quality, multi-terminal electrical contacts on emerging exfoliable and non-exfoliable single-crystal materials for transport measurements, with a focus on methodologies for crystal planarization and metallization. We begin by introducing three key planarization approaches that enable the realization of in-plane multi-terminal electrodes through lithographic processes: (1) mechanical exfoliation of layered crystals to obtain thin flakes, (2) thin lamella lift-out using the focused ion beam (FIB) technique, and (3) bulk crystal planarization via polymeric embedding. For metallization, aside from regular lithographic patterning and metal deposition on planarized surfaces, we highlight several advanced strategies tailored to address the unique challenges posed by specific materials. For instance, dielectric encapsulation and VIA-hole strategies are critical for enabling robust electrical contacts to air-sensitive crystals while mitigating degradation from oxidation or moisture exposure. The micro-stencil strategy offers an alternative route for achieving high-quality contacts by eliminating organic contamination and exposure to excessive heat, solvents, and developers that can otherwise compromise the contacts. Additionally, transfer-based electrode integration methods have shown promise in enhancing contact quality by minimizing interfacial damage and contamination or facilitating interfacial doping. Collectively, these innovations provide valuable insights and practical guidance for the fabrication of electrical contacts to emerging single-crystal systems for reliable transport studies.

## 2. Mechanical exfoliation and contact fabrication of layered crystals

Layered, or exfoliable, crystals represent a major class of materials that have attracted substantial interest in recent transport studies. This is largely attributed to their intrinsic two-dimensional (2D) nature that gives rise to a wide range of unique physical properties, as well as the relative ease of producing lithography-compatible thin flakes, enabling the fabrication of multi-terminal devices tailored for specific functionalities.[68-71] The recent emergence of heterogeneous integration and twist assembly techniques has further broadened the scope for tuning intriguing properties and engineering novel functionalities in these materials.[59,72-74] Notably, efforts to exfoliate and transfer layered transition metal dichalcogenides (TMDs) predate the successful isolation of monolayer graphene in 2004 via the widely known "Scotch tape method".[66] As early as the late 1960s, researchers were already able to prepare various sub-100 nm thick TMD specimens through repeated cleaving using adhesive tapes,[75,76] thereby facilitating subsequent device fabrication using standard microfabrication processes. More recently, Huang et al. introduced a modified mechanical exfoliation method based on a hot cleavage process on $SiO_2$/Si substrates, which significantly improved the yield of exfoliated flakes of graphene and other 2D materials, as shown in Figure 2a and 2b.[77]

On the other hand, as the thickness of 2D crystals is reduced, obtaining sufficiently large thin flakes (typically with lateral dimensions of ~ 50 μm or larger) suitable for multi-terminal electrode configurations becomes increasingly challenging. This limitation is particularly pronounced when targeting the ultra-thin regime or aiming to isolate specific layers for device integration. For example, monolayer or bilayer graphene obtained via regular mechanical exfoliation often exhibits lateral sizes on the order of ~20 μm,[77,78] while the active regions of twisted TMD devices frequently fall below ~ 10 μm.[79,80] Electron-beam lithography (EBL) has become the technique of choice for fabricating such small-scale 2D devices, owing to its sub-micron resolution and high patterning flexibility. Nonetheless, other lithographic approaches, including UV photolithography and direct laser writing, remain viable at the upper limits of resolution capabilities. It is also noteworthy that although millimeter-scale monolayer TMDs have been demonstrated using gold-assisted exfoliation techniques,[82,83] these methods typically require wet etching-based processes to retrieve TMD flakes, which creates additional challenges in preserving the pristine quality of the exfoliated materials.

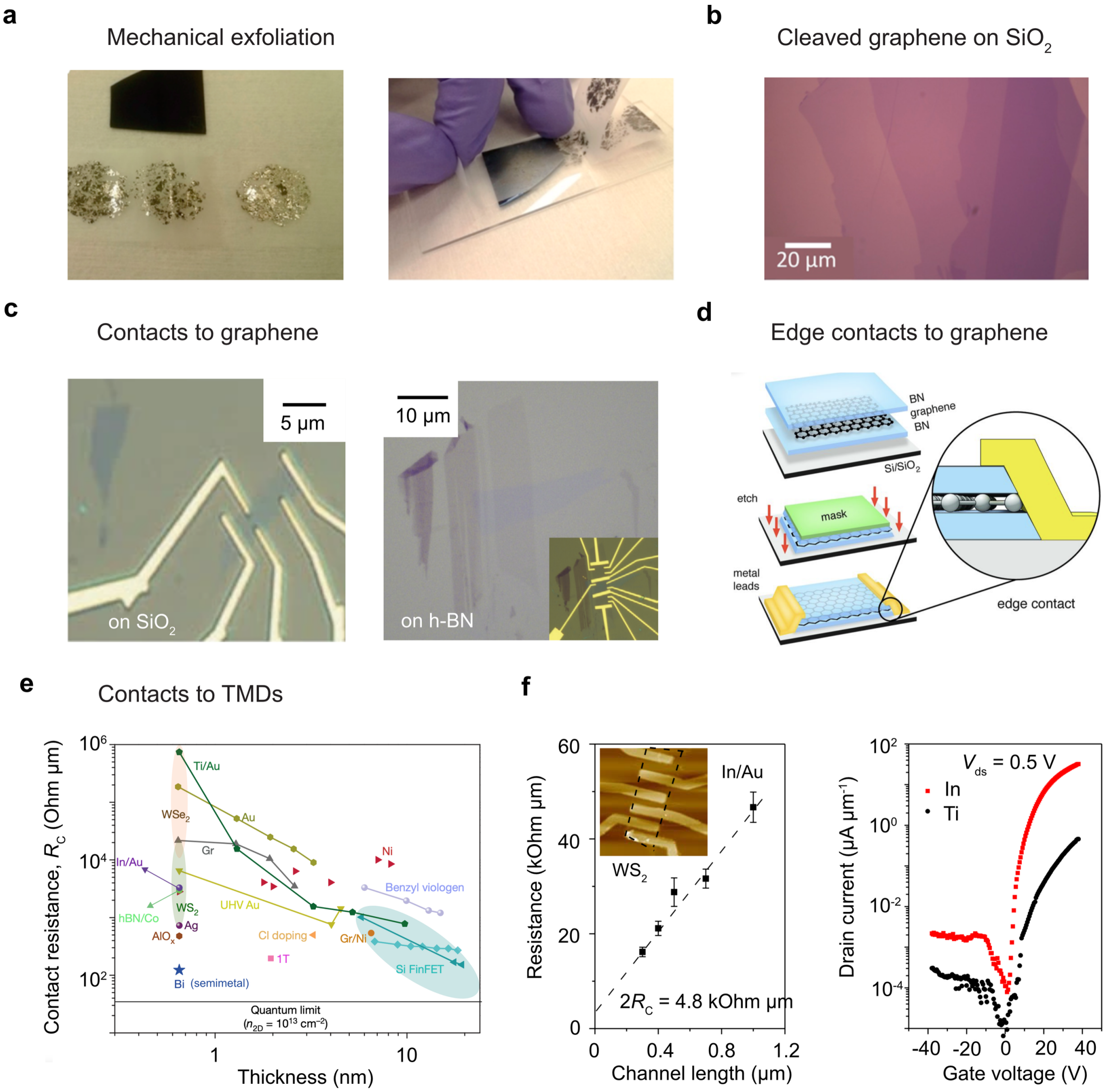

**Figure 2 Contacts to layered crystals**. (**a**) Illustration of the modified mechanical exfoliation process for layered crystals. Figures adapted from Ref. [77] Copyright 2015, ACS Publications. (**b**) Optical micrograph of one representative graphene flake with a thickness varying in steps between 1-4 layers. Figure adapted from Ref. [77] Copyright 2015, ACS Publications. (**c**) Optical micrographs of graphene devices on $SiO_2$ and *h*-BN substrates. Left figure adapted from Ref. [60] Copyright 2005, Springer Nature. Right figure adapted from Ref. [78] Copyright 2010, Springer Nature. (**d**) Schematic illustration of one-dimensional edge contact to graphene. Figure adapted from Ref. [2] Copyright 2013, American Association for the Advancement of Science. (**e**) Scaling of contact resistance $R_C$ with the thickness of $MoS_2$ and ultrathin Si fins or films. The ranges of $R_C$ for monolayer $WS_2$ (shaded green) and $WSe_2$ (shaded orange) are also shown. Figure adapted from Ref. [81] Copyright 2021, Springer Nature. (**f**) Contact properties of In/Au electrodes on mechanically exfoliated $WS_2$, revealed by TLM measurements (left) and transfer characteristics of $WS_2$ transistors (right). Figures adapted from Ref. [53] Copyright 2019, Springer Nature.

Beyond sample size constraints, graphene-based devices are generally considered relatively straightforward to fabricate due to graphene's high electrical conductivity and chemical stability.[84,85] Figure 2c (left) shows an optical micrograph of a graphene device with a Hall-bar geometry, patterned via EBL on thermally grown $SiO_2$. This configuration was used to demonstrate the quantum Hall effect in graphene, as first reported by Zhang et al. in 2005.[60] Subsequently, in 2010, Dean et al. fabricated multiple adjacent graphene devices on both $SiO_2$ and single-crystal hexagonal boron nitride (*h*-BN) substrates (Figure 2c, right), showing that *h*-BN significantly enhances device mobility and suppresses disorder-induced carrier density fluctuations, thereby establishing a gold standard for the fabrication of high-quality 2D electronic devices.[78] Nonetheless, the quality of contacts in graphene devices can still be compromised by polymer residues introduced during flake transfer and stacking processes, often leading to elevated contact resistance. Further complications arise when attempting to electrically access encapsulated 2D layers for metallization. To overcome these limitations, in 2013, Wang et al. developed a novel contact geometry in which the one-dimensional (1D) edge of a 2D graphene sheet is interfaced with a three-dimensional (3D) metal electrode, as illustrated in Figure 2d.[2] This edge-contact scheme enabled remarkably low contact resistance and access to intrinsic electronic properties, including low-temperature ballistic transport and room-temperature mobility approaching the theoretical limit imposed by phonon scattering.[2] In this process, the graphene is first encapsulated between two *h*-BN flakes, after which the entire stack is selectively dry-etched to expose the graphene edges at predefined contact regions for subsequent metallization.[2]

2D semiconductors, particularly TMDs, constitute an important class of layered materials that have attracted attention for their potential applications as post-silicon channel materials in next-generation electronics.[45,71,86] Despite their promising downscaling potential, the performance of field-effect transistors (FETs) based on TMDs is often hindered by inefficient carrier injection, primarily resulting from high contact resistance at the metal-semiconductor interface.[87,88] The recent discovery of correlated electronic phases and superconductivity in twisted TMDs further underscores the pressing need for high-quality electrical contacts in these systems.[74,79,89] Over the past decade, substantial efforts have been devoted to minimizing contact resistance in TMD-based devices, particularly in the ultrathin limit. Several of these strategies are summarized in Figure 2e. For instance, in 2016, English et al. demonstrated that depositing gold (Au) under ultra-high vacuum (~ $10^{-9}$ Torr) significantly reduced the contact resistance to $MoS_2$ by more than threefold compared to conventional deposition conditions.[90] In 2019, Wang et al. reported the realization of ultraclean indium (In) /Au bonding to monolayer $MoS_2$, achieving a record-low contact resistance of ~ 3 kΩ·µm at the time.[53] For mechanically exfoliated few-layer $WS_2$, such In/Au clean contacts yielded contact resistances as low as ~ 2.4 kΩ·µm and significantly improved FET mobility compared to devices utilizing standard Ti/Au electrodes,[53] as illustrated in Figure 2f. Further advancements have explored the use of semimetals such as bismuth (Bi) or antimony (Sb) to suppress metal-induced gap states, enabling ultralow contact resistances of ~ 0.1 kΩ·µm in TMD devices.[81,91] For a more comprehensive overview of contact engineering strategies for 2D TMDs, several informative literature reviews are available elsewhere.[92-94]

It is important to note that the planarization approach based on mechanical exfoliation, when combined with standard lithographic patterning and contact optimization procedures, is broadly applicable to a wide range of layered crystals, provided that two key conditions are met: (1) thin flakes (typically < 100 nm in thickness) with sufficient lateral dimensions (generally > 10 μm) can be readily obtained by mechanical cleavage; and (2) the material is chemically and structurally stable throughout the lithographic and metallization processes. For materials that are prone to degradation upon exposure to air, moisture, elevated temperatures during lithography, or physical damage during metal deposition, specialized metallization strategies must be employed. These advanced techniques, designed to preserve material integrity and ensure reliable contact formation, will be discussed in detail in the second half of this review.

## 3. FIB-based multi-terminal bulk device fabrication

Beyond the layered crystals discussed in the previous section, many interesting material systems such as copper- and iron-based superconductors, as well as arsenide-based semimetals, exhibit 3D or quasi-1D structural characteristics. As a result, these materials cannot be easily cleaved into sub-100 nm thin flakes suitable for regular lithography-based contact fabrication, as is commonly done with 2D materials. Fortunately, significant advances in bulk crystal growth have enabled the synthesis of large-scale single crystals, with lateral dimensions reaching several millimeters or even centimeters.[34,36] This has made it possible to establish electrical contacts on these materials using conventional manual bonding techniques, typically in linear four-probe or six-probe Hall geometries. Despite its success in discovering superconductivity and phase transitions in a wide range of quantum materials from rudimentary transport measurements,[63-65] this method has several severe critical limitations: (1) the substantial thickness of the crystals and the lack of well-defined electrode geometries hinder the accurate extraction of key electronic parameters such as resistivity, carrier concentration, and mobility; (2) difficulties in realizing crystal-orientation-resolved transport measurements; and (3) restrictions in systematic contact optimization.

To overcome these limitations, since the early 2010s, researchers have adopted a powerful and versatile technique known as FIB for the preparation of thin lamellae from bulk crystals and the fabrication of multi-terminal transport devices. Originally, FIB was developed and widely utilized in the semiconductor industry,[95,96] materials science,[97] and biological sciences[98] for site-specific micro-imaging, milling, analysis, and material deposition with exceptional precision. Figure 3a illustrates a typical dual-beam FIB system configuration, in which a scanning electron microscope (SEM) column is combined with an ion beam column to enable precise and localized processes in a single platform.[97] In physical sciences, FIB has traditionally been employed for the preparation of ultrathin lamellae (< 100 nm thick) required for transmission electron microscopy (TEM) analysis.[99,100] The application of FIB in transport device fabrication builds upon similar procedures. As shown in Figure 3b, a thin lamella, typically several micrometers thick, is first defined and carved from a bulk crystal by ion milling (left), then extracted using a micromanipulator and lifted out from the parent crystal (right).[99] While both TEM sample preparation and transport device fabrication use the same lift-out procedure, the downstream processes differ: for TEM, the lamella is attached to a TEM grid and further thinned down to below 100 nm; in contrast, for transport device fabrication, the lamella is mounted horizontally onto a substrate using epoxy, followed by metallization and localized FIB milling to define desired contact geometries.

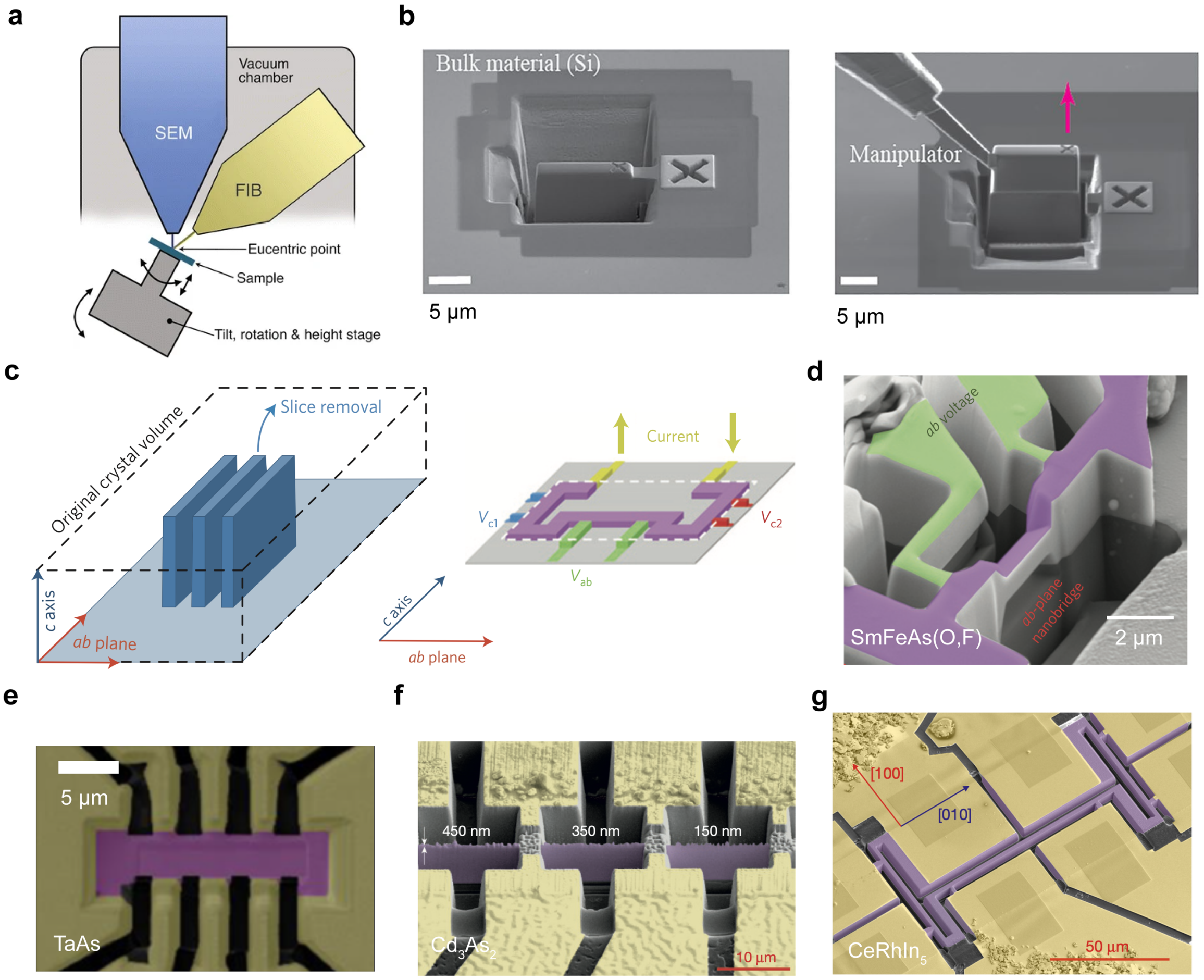

**Figure 3 FIB fabrication**. (**a**) Schematic illustration of a dual-beam FIB/SEM setup. Figure adapted from Ref. [97] Copyright 2014, Springer Nature. (**b**) SEM images of FIB lift-out processes for preparing a lamella. Figures adapted from Ref. [99] Copyright 2021, Springer Nature. (**c**) Schematic illustration of lamella cutting (left) and U-shape micropatterning (right) processes of a SmFeAs(O, F) crystal. Figures adapted from Ref. [101] Copyright 2010, Springer Nature. (**d**) False-colored SEM image of a FIB-fabricated nanobridge of SmFeAs(O, F) with cross-sectional dimensions of 600 nm × 600 nm. Figure adapted from Ref. [101] Copyright 2010, Springer Nature. (**e**) to (**g**) SEM images of FIB-prepared transport devices from various newly developed crystals, including TaAs, $Cd_3As_2$, and $CeRhIn_5$. Figures adapted from Ref. [102] Copyright 2019 Springer Nature, Ref. [103] Copyright 2016, Springer Nature, and Ref. [104] Copyright 2017, Springer Nature.

In a pioneering study, Moll et al. utilized Ga-FIB milling to extract an oriented lamella from a bulk crystal of the iron-based superconductor SmFeAs(O, F), which was subsequently patterned into three continuous four-probe resistance bars, one in the *ab* plane and the other two along the *c*-axis, as illustrated in Figure 3c.[101] By injecting current through the outer electrodes and measuring voltage drops across the inner leads ($V_{c1}$, $V_{c2}$, and $V_{ab}$), the resistivity along different crystallographic orientations could be simultaneously probed. This configuration proved particularly advantageous for investigating electronic anisotropies within the same device. Furthermore, this FIB-carving approach enabled direct measurements of high critical current densities of ~ $1\times10^6$ A/cm$^2$ by confining current flow through a freestanding FIB-fabricated nanobridge of SmFeAs(O, F), with cross-sectional dimensions of approximately 600 nm × 600 nm, as shown in Figure 3d.[101] Building upon this methodology, Moll et al. expanded the application of FIB-based transport device fabrication to a variety of interesting single-crystal systems, including the Weyl semimetal TaAs (Figure 3e),[102] the Dirac semimetal $Cd_3As_2$ (Figure 3f),[103] and the heavy-fermion superconductor $CeRhIn_5$ (Figure 3g).[104] More recently, FIB milling has been increasingly adopted by research groups worldwide, establishing it as a versatile and powerful tool for creating high-quality transport devices from emerging single crystals.[105-108]

In contrast to the conventional manual bonding approach, which places stringent constraints on crystal dimensions and geometries, FIB-based micropatterning offers several key advantages. First, it enables the fabrication of multi-terminal transport devices from sub-millimeter, non-exfoliable crystals, as the required lamellae typically measure less than 100 μm laterally and only a few microns in thickness. Second, the substantially reduced sample thickness and optimized device geometry enhance the magnitude of measurable signals such as resistances and Hall voltages, thereby rendering previously indiscernible transport features experimentally resolvable. Third, FIB processing allows lamellae to be prepared along specific crystallographic orientations and carved into desired device geometries, including Hall bars and angle-resolved resistance bars, thus providing an ideal platform for probing electronic anisotropies.

Despite its clear advantages, the widespread adoption of FIB-based fabrication for transport devices remains limited by several practical challenges. These include the high cost of instrumentation, operational complexities, and limited machine time in many research laboratories. Moreover, the use of high-energy ion beams inevitably induces structural defects or unintentional doping in the vicinity of the crystal surface, which can lead to extrinsic modulation of the material's transport characteristics.[109,110] Although post-processing steps such as low-kV ion polishing or gentle argon-ion milling can partially mitigate this damage, they may not entirely eliminate it. Additionally, the FIB fabrication process often induces non-uniform strain fields within the micro-structured devices,[111] potentially complicating the interpretation of transport measurements by superimposing strain-related effects on intrinsic transport behavior. This concern is particularly relevant for materials with strong electron-electron correlations or significant electron-phonon coupling. Therefore, when employing FIB-fabricated devices for transport studies, it is crucial to perform complementary characterization, such as transport measurements using conventionally prepared, damage-free devices, or structural analysis, to ensure that the observed phenomena accurately reflect the intrinsic properties of the material.

## 4. Polymeric planarization strategies for small, non-exfoliable crystals

For many years, lithography-based contact fabrication techniques have been primarily restricted to wafer-scale samples or thin flakes (usually < 100 nm) exfoliated onto flat substrates, where uniform coverage of photoresist or electron-beam resist prepared by spin-coating can be reliably achieved. In contrast, most non-exfoliable crystals synthesized through laboratory-scale techniques such as CVT or flux growth without extensive optimization possess small lateral dimensions but substantial thicknesses (often tens of micrometers or larger), making it extremely challenging to achieve continuous and uniform resist coverage.[34,36] Consequently, although FIB-based fabrication techniques offer a viable route for thinning and metallizing such crystals, there remains a strong demand for alternative crystal planarization strategies that are compatible with standard lithographic processes.

Different from top-down crystal thinning approaches such as mechanical exfoliation commonly employed for layered materials, the polymeric planarization method discussed in this section represents a bottom-up strategy. This technique involves filling the gap between the substrate and the top surface of a small, bulk crystal using a polymeric matrix, thereby enabling a flat, uniform surface compatible with standard microfabrication processes. As illustrated in Figure 4a, a microscale crystal or device is first mounted onto a polydimethylsiloxane (PDMS) stamp and then immersed in a drop of UV-curable adhesive (e.g., NOA 61, Norland). Once the adhesive has flowed to completely fill the gap between the PDMS and substrate, UV exposure is used to cure the epoxy.[112,113] The PDMS is then carefully removed, leaving a planarized crystal surface suitable for lithography-based contact fabrication and optimization.[112,113]

This concept of polymeric embedding for handling small crystals was originally developed for applications such as microtome sectioning[114,115] and surface polishing for chemical analysis.[116] In the late 2010s, Yoon and collaborators adopted this approach to fabricate lithographically defined contacts on GaAs or Si microcells (typically 5-10 μm thick) that were planarized using UV-curable epoxy following epitaxial lift-off (ELO), which has enabled the development of several high-efficiency photovoltaic devices[117-119] and durable III-V photoelectrodes for solar water splitting.[113] In 2024, Zhao et al. successfully applied the NOA-based embedding method to demonstrate the first single-crystal based photoconductive device of the perovskite chalcogenide $BaZrS_3$, using crystals with lateral dimensions of 100-200 μm and thicknesses exceeding 100 μm.[23] A cross-sectional schematic illustration of the device is shown in Figure 4b. It is also worth noting that a gentle mechanical polishing process was employed in this prototype device prior to contact fabrication to freshly expose the crystal surface. While this step enabled successful electrode integration, it also introduced a significant density of surface defects, which adversely impacted device performance.[23] The resulting two-terminal photoconductive device exhibited relatively slow responses, with rise and decay times of 18 s and 26 s, respectively.[23] An optical micrograph of the device is presented in Figure 4e.

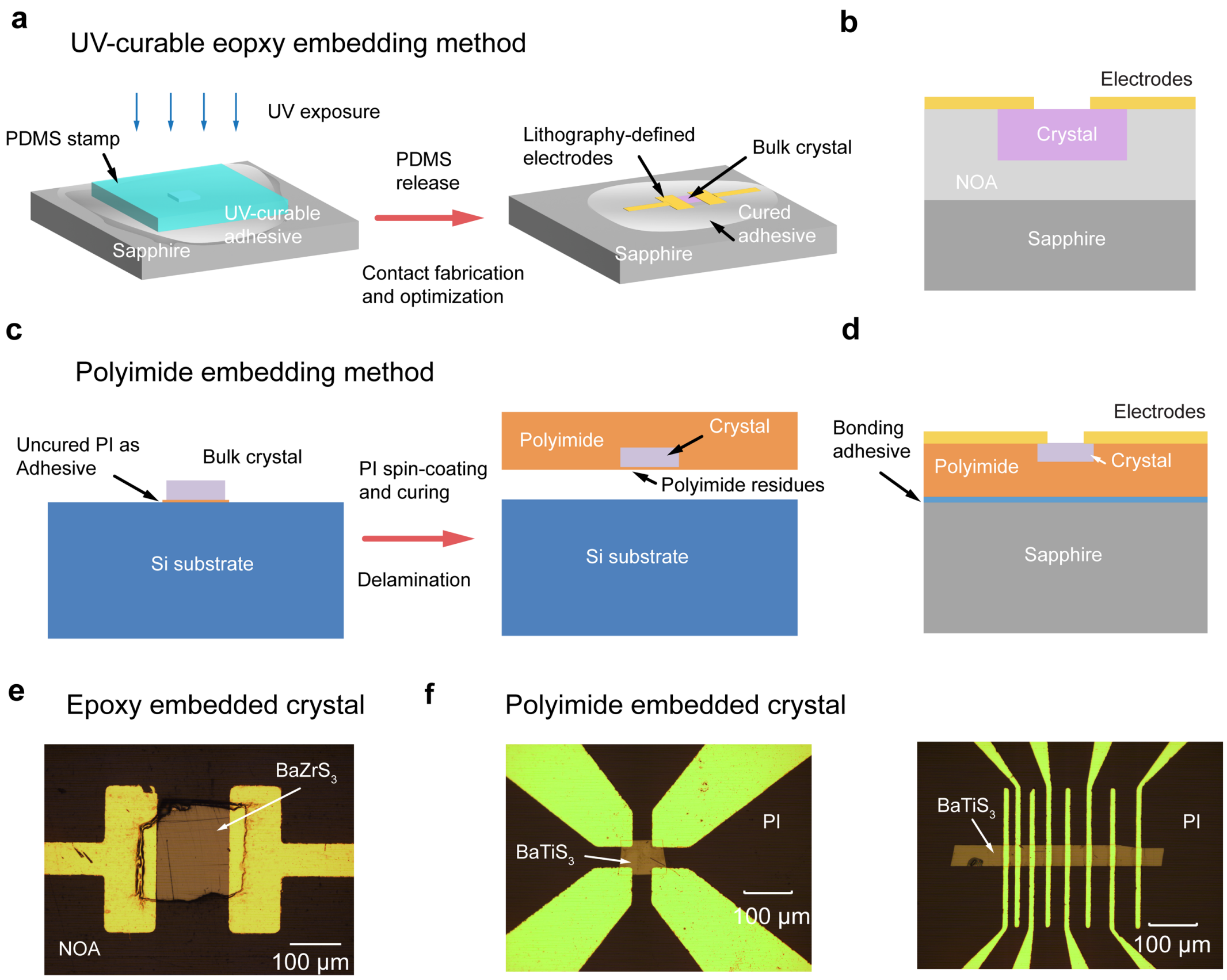

**Figure 4 Polymeric planarization strategies**. (**a**) Schematic illustration of planarization process using UV-curable epoxy. Figures adapted from Ref. [112] Copyright 2025, Springer Nature. (**b**) Cross-sectional illustration of a single-crystal device fabricated using epoxy-based planarization method. (**c**) Schematic illustration of fabrication processes for crystal planarization using polyimide. Figure adapted from Ref. [108] Copyright 2022 ACS Publications. (**d**) Cross-sectional illustration of a polyimide-planarized device. (**e**) Optical micrographs of a $BaZrS_3$ device prepared by NOA embedding. (**f**) Optical micrographs of $BaTiS_3$ devices with contact geometries of van der Pauw (left) and TLM (right), prepared through polyimide planarization and encapsulation. Figures adapted from Ref. [6] Copyright 2023 John Wiley & Sons.

While epoxy embedding offers a straightforward and effective route for planarizing small crystals for lithographic processing, most commercially available epoxies exhibit relatively high coefficients of thermal expansion (CTE). For example, NOA 61 possesses a CTE of ~80 ppm/K, which can induce pronounced thermal strain, on the order of ~ 1.6% at 100 K, that may compromise the integrity of low-temperature transport measurements. In 2023, Chen et al. reported inconsistent transport behavior in a phase-change $BaTiS_3$ crystal embedded in NOA epoxy, where the transition temperatures were either progressively shifted or entirely suppressed after repeated thermal cycling due to extrinsically induced thermal strain.[6] It is also worth noting that, under certain circumstances, differential thermal expansion between the sample and its bonding substrate can be intentionally exploited to introduce in-plane biaxial strain, as demonstrated in single crystals of $Ca(Fe_{1-x}Co_x)_2As_2$.[120]

To mitigate thermal strain effects in planarized crystals such as $BaTiS_3$, Chen et al. developed a low-stress polymeric planarization technique, as illustrated in Figure 4c.[6,108,121] Although the CTE of $BaTiS_3$ is not reported in the literature, a polyimide (PI 2611, HD Microsystems) with a low CTE of ~ 3 ppm/K was selected as the embedding medium, based on the general observation that most inorganic single crystals exhibit CTE values below 10 ppm/K. In this approach, a $BaTiS_3$ crystal with a thickness of ~ 20 μm is first identified under an optical microscope and temporarily bonded to a Si carrier substrate using the partially cured polyimide as adhesive. This is followed by multiple spin-coating steps (four times) to produce a ~ 60 μm-thick polyimide film that embeds the crystal.[108] Crucially, no silane-based adhesion promoters are used during this process, allowing the fully cured polyimide film to be cleanly delaminated from the Si carrier and subsequently bonded to a sapphire substrate for electrode fabrication after flipping. As a result of the planarization, the step height at the $BaTiS_3$ crystal-substrate boundary is reduced from the initial crystal thickness (5-20 μm) to less than 200 nm, making the surface fully compatible with lithographic processing.[108] It is important to note, however, that due to limitations inherent to the spin-coating and thermal imidization processes, this polyimide-embedding approach has not yet been successfully extended to crystals thicker than 30 μm. To extend its applicability to thicker samples, specially designed casting and drying procedures for forming thicker polyimide films may be required, which fortunately are already well established in the polyimide manufacturing industry.[122,123]

Following polyimide-based planarization, the subsequent processes largely follow standard contact fabrication procedures used for thin films, 2D samples, or epoxy-embedded crystals. The primary distinction in the polyimide-based approach is the inclusion of an additional, gentle reactive ion etching (RIE) step to remove the residual polyimide and freshly expose the crystal surface prior to metallization.[6,108] A cross-sectional schematic of a polyimide-embedded $BaTiS_3$ device is shown in Figure 4d, and representative optical micrographs of fabricated devices with varying contact geometries are presented in Figure 4f. Notably, transport measurements performed on $BaTiS_3$ devices fabricated using this polyimide planarization approach consistently revealed two hysteretic phase transitions at cryogenic temperatures, observed across multiple devices and thermal cycles.[6] These transitions are in agreement with both structural analysis and the transport behavior of reference devices fabricated via manual bonding. The reproducibility and consistency of these results affirm the effectiveness of the polyimide embedding strategy in mitigating strain-related effects and preserving the intrinsic transport properties of strain-sensitive crystals such as $BaTiS_3$.[6]

In summary, both the epoxy-based embedding method and the low-stress polyimide planarization strategy offer effective solutions for planarizing small, non-exfoliable crystals to enable lithography-compatible fabrication of multi-terminal electrodes, which has been a long-standing challenge in transport studies of emerging materials. The epoxy-based approach is particularly suitable for preparing devices that operate at room temperature and for quick preliminary transport tests, owing to its simplicity and ease of implementation, whereas for handling strain-sensitive systems such as phase-change materials, the low-stress polyimide planarization method is preferred to preserve their intrinsic transport properties. Additionally, other UV- or thermally curable epoxies may also be adapted for crystal planarization following procedures analogous to those described above, offering further flexibility in accommodating a wide range of materials and device requirements.

**Table 1 Comparison of planarization methods for contact fabrication on emerging single crystals.**

| Planarization methods | | Key advantages | Principal limitations | Precautions | Representative examples |
|---|---|---|---|---|---|
| Mechanical exfoliation | | Simple process; lithography-ready flakes; compatible with *h*-BN and vdW heterostructures | Thin flakes are small in size; often require EBL for patterning | Minimize transfer residue; thin air-sensitive flakes degrade faster | Graphene [60], $WS_2$ [53], CeSiI [18] |
| FIB lift-out | | Oriented lamellae and μm-scale geometries (Hall bars, nanobridges); enhanced signals; metallization built in | Costly, with limited access; risks of surface damage, Ga-doping and non-uniform strain | Finish with a low-kV polish; verify against damage-free devices, especially for correlated systems | SmFeAs(O,F) [101], TaAs [102], $CeRhIn_5$ [104] |
| Polymeric embedding | Epoxy | Simple process; handles thick crystals | Introduces substantial thermal strain at cryogenic temperatures due to the large CTE of epoxy | Limited to room-temperature devices and quick tests | $BaZrS_3$ [112]; strain artifacts in $BaTiS_3$ [6] |
| | Low-stress polyimide | Preserves intrinsic transport properties; enables encapsulation | Only crystals < 30 μm thick so far; extra curing and RIE steps | Release polyimide film from Si wafer; use a gentle RIE to expose crystal surface | $BaTiS_3$ [6] |

## 5. Dielectric encapsulation strategies for handling 'sensitive' crystals

In the preceding sections, we reviewed three key planarization strategies, i.e., mechanical exfoliation, FIB thinning, and polymeric embedding, which have enabled multi-terminal contact fabrication and transport studies on many emerging crystals featuring exciting physical properties. However, a significant subset of materials is highly sensitive to ambient conditions, exhibiting rapid degradation upon exposure to air or moisture, making them incompatible with conventional microfabrication processes. Notable examples include many alkali- and alkaline-earth-containing compounds (e.g., LiFeAs, $BaNi_2As_2$),[124,125] halides (e.g., $CrI_3$, CeSiI),[18,126] and chalcogenides (e.g., $MoTe_2$, $NbSe_2$).[127,128] The degradation of such materials is often exacerbated upon heating or in the thin limit, necessitating the use of dielectric encapsulation layers to protect pristine crystal surfaces from oxygen and moisture. To electrically access the encapsulated crystals, lithographically defined vertical interconnect access (VIA) holes can be employed, allowing for precise electrode integration without exposing the sensitive materials. In this section, encapsulation strategies for both mechanically exfoliated 2D flakes and polymer-planarized non-exfoliable crystals are discussed.

Hexagonal boron nitride (*h*-BN), one of the few layered insulators with atomically flat and chemically inert surfaces, plays a pivotal role in the fabrication and characterization of 2D materials. It is widely employed as a high-quality insulating substrate and gate dielectric.[78,129] More critically, *h*-BN serves as an effective encapsulation layer for air-sensitive 2D materials, particularly in the ultrathin regime, to preserve their intrinsic physical properties.[18,126] Figure 5a shows an example in which a thin *h*-BN flake was placed onto an air-sensitive CeSiI device after mechanical exfoliation in an inert atmosphere glovebox and contact fabrication through a shadow mask.[18] Such a capping strategy is highly effective in preserving sample integrity for optical and X-ray-based characterizations. However, for transport measurements, electrical contacts need to be additionally established to access the encapsulated active layers. This is typically achieved by creating 'VIAs' through the *h*-BN layer for contacts using lithography and RIE, as shown in Figure 5b. It is important to note that when the etching selectivity is poor, both the *h*-BN layer and the underlying 2D device can be simultaneously etched, leading to direct contact formation at the exposed edge of the 2D crystal, as in the case of 1*T*-$TaS_2$ flakes.[130] In the same study, Tsen et al. demonstrated the effectiveness of *h*-BN encapsulation in preserving phase transitions in ultrathin 1*T*-$TaS_2$ flakes.[130] Specifically, they observed the nearly commensurate to commensurate charge density wave (NCCDW-CCDW) transition in samples as thin as 4 nm, a phenomenon previously believed to be suppressed in flakes thinner than 10 nm.[131]

For polymer-planarized, non-exfoliable crystals, polyimide-based encapsulation combined with lithographically defined VIA structures offers a promising route for handling air-sensitive bulk crystals. Building upon the polyimide-embedding method discussed in Section 4, an additional thin polyimide layer is spin-coated and thermally cured on the planarized crystal surface to achieve full encapsulation. VIA holes are then lithographically patterned and dry-etched through the polyimide layer to freshly expose the underlying crystal surface for electrical contact formation, as illustrated in Figure 5c.[6,132] As one of the most widely adopted encapsulation materials in semiconductor and flexible electronics industries,[133,134] the processing techniques of polyimide, including spin-coating, curing, dry etching, and subsequent electrode integration, are well-established and documented in the literature.[122,135,136] It is also worth noting that these methods were initially developed to enable precise definition of contact regions and Hall bar geometries in $BaTiS_3$ crystals, facilitating accurate extraction of intrinsic transport parameters such as resistivity, carrier concentration, and mobility.[6] Figure 5d shows two representative microfabricated $BaTiS_3$ devices (100-200 μm laterally) featuring standard and orientation-resolved multi-terminal Hall bar configurations.[132,137] In addition to geometric precision, the polyimide-based VIA approach offers unique advantages for preserving the integrity of crystals susceptible to degradation. By effectively shielding the embedded crystal from ambient oxygen and moisture, the polyimide dielectric functions analogously to *h*-BN encapsulation in 2D materials, thereby enabling reliable and reproducible transport measurements in otherwise unstable systems.

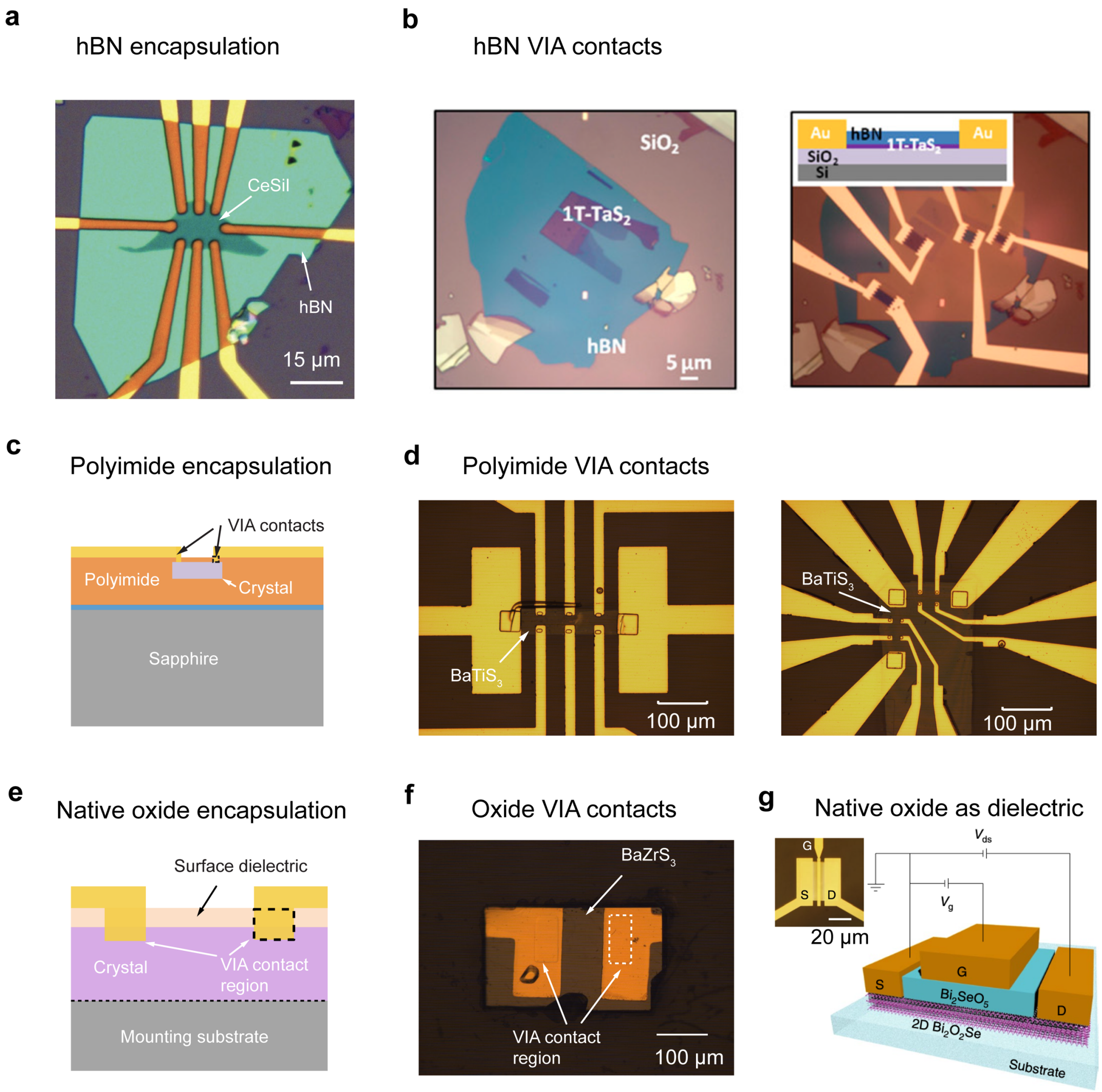

**Figure 5 Dielectric encapsulation and VIA strategies**. (**a**) Optical micrograph of a CeSiI transport device encapsulated by *h*-BN. Figure adapted from Ref. [18] Copyright 2024, Springer Nature. (**b**) Optical microscopic images of 1*T*-$TaS_2$ flakes on a $SiO_2$/Si wafer covered by *h*-BN before (left) and after (right) side electrical contact. Side-view device schematic is shown as the inset. Figures adapted from Ref. [130] Copyright 2015, National Academy of Sciences. (**c**) Cross-sectional illustration of a polyimide-planarized device contacted through VIA holes. (**d**) Optical micrographs of $BaTiS_3$ devices with contact geometries of standard (left) and orientation-resolved Hall bars (right), prepared through polyimide planarization and VIA hole method. Left figure adapted from Ref. [132] Copyright 2024, John Wiley & Sons. Right figure adapted from Ref. [137] Copyright 2023, Huandong Chen. (**e**) Cross-sectional illustration of a single-crystal device contacted through native oxide VIA holes. (**f**) Optical micrograph of a single-crystal $BaZrS_3$ device contacted through oxide VIAs. Figure adapted from Ref. [112] Copyright 2025, Springer Nature. (**g**) Schematic illustration of a $Bi_2O_2Se$ transistor with $Bi_2SeO_5$ as the gate dielectric. The inset shows an optical micrograph of the device. Figures adapted from Ref. [138] Copyright 2020, Springer Nature.

Certain air-sensitive crystals exhibit a tendency to form surface oxide layers upon exposure to oxygen or moisture, rather than undergoing direct decomposition.[139,140] These dense oxide layers act as passivating barriers, effectively inhibiting further diffusion of oxygen into the underling materials. As such, they can be leveraged as natural dielectric capping layers for air-sensitive materials. Similar to other encapsulation strategies, electrical contacts can be established through lithographically defined VIA holes, created by local dry etching of the surface dielectric, as illustrated in Figure 5e. A notable example is the photoactive semiconducting crystal $BaZrS_3$, which presents significant challenges for achieving high-quality electrical contacts due to (1) the formation of sulfate or sulfide layers during the DI water rinsing step following flux growth, and (2) the oxidation propensity of zirconium species under ambient or thermal processing conditions.[23,112] In the first reported $BaZrS_3$ single-crystal photoconductive device (2024), an epoxy-planarized crystal was mechanically polished using fine sandpaper to remove surface dielectrics prior to metallization. However, this brute-force approach introduced a high density of surface defects, resulting in substantial dark current and slow photoresponse characteristics ($\tau_{rise}$ = 18 s, $\tau_{decay}$ = 26 s), attributed to carrier trapping at defect sites.[23] More recently, Chen et al. developed a dry-etching-based, polishing-free contact fabrication procedure for planarized $BaZrS_3$ crystals (Figure 5f), in which only the VIA contact regions are exposed to dry etching, leaving the channel region intact. This oxide dielectric encapsulation strategy yielded a two-orders-of-magnitude reduction in dark current and significantly improved switching performance ($\tau_{rise}$ = 0.18 s, $\tau_{decay}$ = 0.2 s).[112] Alternatively, researchers have also intentionally induced surface oxidation and used the native oxide layer as a gate dielectric, as illustrated in Figure 5g.[138]

In summary, the dielectric encapsulation strategies discussed above are essential for enabling reliable transport measurements for various air-sensitive crystals. Specifically, the pick-and-drop-based *h*-BN encapsulation method is particularly well-suited for 2D devices, while the spin-coating-based polyimide strategy is more compatible with polymer-planarized bulk crystals. For FIB-fabricated devices, ion-beam-induced damage or degraded surfaces in the channel regions are typically mitigated by a final low-voltage milling step. Following this, additional encapsulation using polyimide or other low-stress dielectric materials can further protect the device and help preserve the intrinsic electronic properties of the underlying crystal.

## 6. Micro-patterned shadow mask strategy for contact fabrication

In recent years, lithography-based microfabrication techniques have been increasingly adopted for contact fabrication on a wide range of emerging crystals, often in conjunction with crystal planarization strategies such as mechanical exfoliation or polymeric embedding. However, both photolithography and electron beam lithography expose crystals to potentially detrimental processing conditions, including elevated temperatures during soft- and/or post-exposure bakes, aqueous solutions during resist development, and polymer residues. As a result, these processes cannot be directly employed to handle air-sensitive crystals without encapsulation or to create contacts for applications requiring pristine surfaces, such as scanning tunneling microscopy (STM) or conductive atomic force microscopy (c-AFM). An effective alternative involves the use of micro-patterned shadow masks for metal deposition (Figure 6b). This approach eliminates the need for thermal processing, chemical treatment, or, in certain cases, even exposure to polymers, while still enabling reasonably high precision in defining electrode geometries. As such, microstencil-based deposition is particularly well-suited for small, air-sensitive crystals or microflakes. In comparison to macroscopic shadow masks (e.g., tapes, metal strips or stencils) that are commonly used for stencil printing or contacting large crystals, the primary limitation of micro-scale shadow mask techniques lies in the relative complexity of mask fabrication. Commonly used materials for microstencils include silicon nitride ($SiN_x$) membranes, thin silicon (Si) wafers, and polymer films, each of which will be discussed in detail below.

Figure 6a illustrates the representative fabrication process for $SiN_x$-based shadow masks. The procedure begins with the deposition of several hundred nanometers of low-stress $SiN_x$ on both sides of a (001)-oriented Si wafer using low-pressure chemical vapor deposition (LPCVD). Anisotropic wet etching of the underlying Si is then performed from the backside through a lithographically patterned $SiN_x$ window using hot potassium hydroxide (KOH).[141] The use of low-stress $SiN_x$ films is essential to achieving self-supporting freestanding membranes after the Si etch.[142] Notably, such LPCVD-grown $SiN_x$ wafers, as well as $SiN_x$ membranes, are also commercially available owing to their high demand in versatile microelectromechanical systems (MEMS) applications. The desired electrode patterns are then lithographically defined and carved into the $SiN_x$ membrane by reactive ion etching (RIE).[143] Due to the fragile nature of these ultrathin membranes, special care must be taken during processing to avoid film cracking. For instance, thin layers of polymethyl methacrylate (PMMA) resists are generally not preferred due to their limited resistance to dry etching, which may lead to insufficient protection of the membranes; in contrast, many standard photoresists and other electron beam (e-beam) resists (e.g., ZEP series) would work well for this process. Although $SiN_x$ shadow masks themselves are reusable, an oxygen plasma-based cleaning step is typically employed to refresh the mask surface before each use, which in turn gradually thins down the $SiN_x$ membrane upon repeated plasma exposure, thereby limiting its overall lifetime.

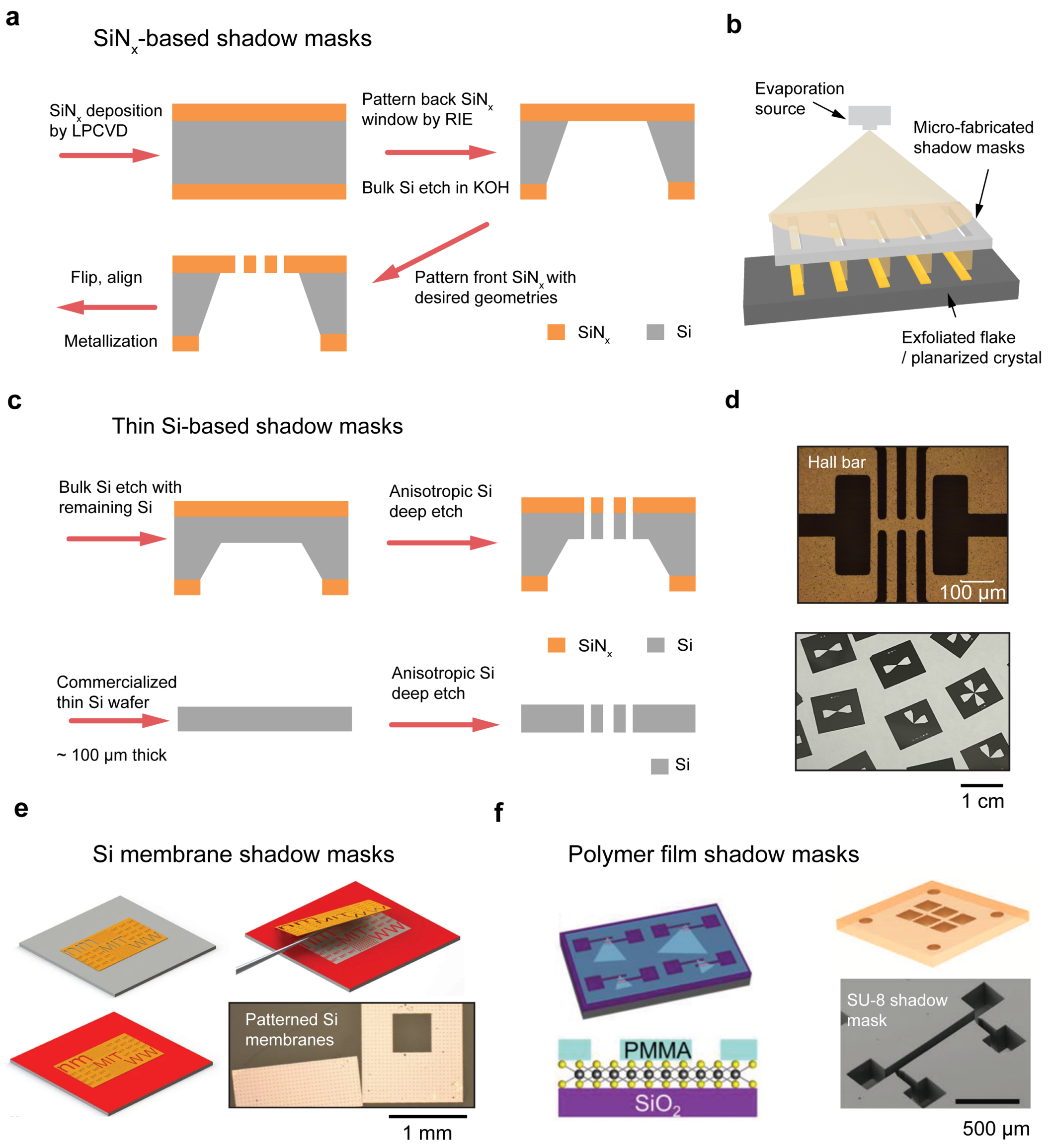

**Figure 6 Micro-patterned shadow mask strategy**. (**a**) Schematic illustration of fabrication processes for a $SiN_x$-based shadow mask. (**b**) Illustration of contact fabrication process through a shadow mask. (**c**) Schematic illustration of fabrication processes of thin Si-based shadow masks using $SiN_x$/Si wafers (top) and thin Si wafers (bottom). (**d**) Optical images of thin Si-based shadow masks with geometries of a Hall bar (top) and various two-terminal and four-terminal patterns (bottom). (**e**) Schematic illustration of metal deposition process using a Si membrane shadow mask. The bottom right shows an optical image of patterned Si membranes on a quartz substrate. Figures adapted from Ref. [144] Copyright 2015, Springer Nature. (**f**) Illustration of polymer-based shadow masks fabricated from PMMA (left) and SU-8 (right). PMMA, Ref. [145] Copyright 2021, John Wiley & Sons. SU-8, Ref. [146] Copyright 2019, John Wiley & Sons.

Alternatively, one may seek to use thin Si-based shadow masks for improved mechanical robustness and durability.[147] As illustrated in Figure 6c (top), such masks can be fabricated following a similar process flow, except that a portion of the bulk Si (typically 50-100 μm thick) is intentionally preserved during the KOH-based anisotropic wet etching step.[141] The desired microstencil patterns are then engraved through subsequent $SiN_x$ dry etching and deep Si etching, using cryogenic or Bosch processes, with hard metal masks such as chromium or nickel films.[148,149] Directly deep-etching commercially available thin Si wafers (50-100 μm thick) is also a viable solution, as shown in Figure 6c (bottom), provided that such thin wafers can be handled properly without breakage. Figure 6d presents two optical images of Si-based shadow masks fabricated by the authors: one featuring a Hall bar geometry and the other containing a variety of two-terminal and four-terminal patterns. Compared to $SiN_x$-based shadow masks, Si microstencils exhibit superior mechanical robustness, greater resistance to repeated surface cleaning (e.g., oxygen plasma treatments), and significantly improved reusability, at the cost of slightly reduced pattern resolution. For example, a 100 μm-thick Si shadow mask typically supports minimum feature sizes larger than 2 μm even if the dry-etch process is well optimized for Si deep etch to achieve a 50:1 aspect ratio, while for membranes (~ 300 nm thick), sub-micron features are easily achievable. It is also worth noting that for both Si- and $SiN_x$-based rigid stencils, a microgap between the shadow mask and the substrate is often unavoidable during integration, due to the rigidity of the mask and unevenness of the substrate over millimeter length scales. This gap can cause metal spreading issue during deposition that largely limits the final pattern resolution, particularly in the sub-micron resolution regime.

Other candidates include lithographically patterned flexible Si membranes and various polymer films, both of which can be integrated conformally onto the substrate for metal deposition. Each strategy offers distinct advantages and limitations. For instance, ultrathin Si membranes (~ 220 nm thick), released from commercially available silicon-on-insulator (SOI) wafers, have enabled the fabrication of a shadow mask with spatial patterning resolution down to 10 nm (Figure 6e).[144] Such superior pattern transfer resolution of a shadow mask is achieved by combining (1) excellent lithography resolution from optimized EBL processes, (2) negligible resolution loss from dry etching due to the ultrathin thickness, and (3) no additional resolution loss from the deposition step owning to conformal integration of the mask. On the other hand, unlike previously discussed wafer-supported $SiN_x$ or Si masks, these freestanding Si membrane stencils require careful handling *via* transfer-based techniques and are generally not reusable, despite their exceptional patterning resolution.

As for the polymer film-based shadow masks, fabricated from organic resist materials such as PMMA or SU-8 (Figure 6f),[145,146] their advantages lie in the process simplicity and low cost. The resist etching process is done within the lithography step through developing, and hence, no additional dry etch is required before releasing the polymer mask for integration. However, the inevitable conformal contact between the polymer stencil and the active material and the resulting organic residues can render this method unsuitable for preparing contacts for surface-sensitive characterizations such as STM. Nonetheless, compared to conventional lithographic processes using spin-coated resists, the polymer film shadow mask strategy avoids exposing the contact regions to high-energy radiation (e.g., electron beam or UV), excessive heat, and chemicals. As a result, improved performance has been demonstrated recently in 2D transistors fabricated using the transferred polymer stencil approach,[145] while inorganic material-based shadow masks using Si or $SiN_x$ are still among the best options to prepare contacts of ultraclean 2D devices for STM measurements.[150,151]

Overall, while micro-patterned shadow masks offer several compelling advantages for fabricating multi-terminal contacts, particularly for air-sensitive materials or surface-sensitive techniques, several practical limitations remain. For instance, the precise alignment of shadow masks with small crystals or flakes can be nontrivial. Typically, alignment is performed manually under an optical microscope, with the mask position secured using adhesive tapes at the corners or a small amount of vacuum grease, which is particularly challenging for microscale 2D flakes. Moreover, the small gap between these rigid shadow masks and the crystal surface largely limits patterning accuracy and precludes the use of conformal deposition-based techniques such as sputtering. It is also worth noting that although both ultrathin Si membrane and polymer-based shadow masks can achieve excellent pattern resolution and form more conformal contact with crystals, their poor durability and potential surface contamination of crystals (specifically for polymer-based shadow masks) have further limited their widespread application.

## 7. Transfer-based electrode integration for improved contacts

Establishing high-quality electrical contacts to emerging materials, particularly gapped systems, is essential for probing their intrinsic electronic properties and for maximizing the performance of associated electronic and optoelectronic devices. In well-established semiconductors such as Si and GaAs, heavily doped contact regions are often intentionally introduced to facilitate the formation of low-resistance Ohmic contacts,[152-154] while for many newly synthesized crystals, achieving reliable contacts remains challenging. In principle, Schottky barrier heights at the metal-semiconductor interface can be minimized by selecting contact metals with work functions aligned to the conduction or valence band edges of the material.[155] However, such ideal alignment is often not realized due to variations in the interfacial structure. Specifically, interface damages caused by large kinetic energies of the deposited species during metal deposition can lead to substantial Schottky barriers and/or large contact resistance.[67,156,157] In this section, we use semiconducting TMDs as examples and review recent developments of device fabrication strategies, primarily transfer-based, to improve contact quality. Various effective ways to reduce contact resistance in TMDs through optimizing metal deposition processes have been discussed previously in Section 2.

A promising strategy to improve electrical contacts to TMDs involves the transfer of pre-patterned metal electrodes, rather than direct metal deposition. In 2018, Liu et al. demonstrated the formation of van der Waals (vdW) metal-semiconductor junctions by laminating atomically flat metal electrodes onto the clean, dangling-bond-free surfaces of TMDs, thereby circumventing the chemical disorder and defect-induced gap states commonly introduced by metal evaporation processes, as illustrated in Figure 7a.[67] Notably, for these transferred contacts, the Schottky barrier heights were observed to follow the Schottky-Mott law, exhibiting strong dependence on the metal work function. The behavior contrasts sharply with evaporated metal contacts, which typically show weak or no correlation due to pronounced Fermi-level pinning at the metal-semiconductor interface.[67,158] Using this approach, a prototype asymmetric Ag-$MoS_2$-Pt diode exhibiting appreciable photovoltaic characteristics ($V_{OC}$ ~ 1.02 V) was demonstrated.[67] The same strategy was later employed to fabricate contacts on single-crystalline thin films of $CsPbBr_3$, enabling detailed investigations of its photoelectrical transport phenomena at cryogenic temperatures.[159]

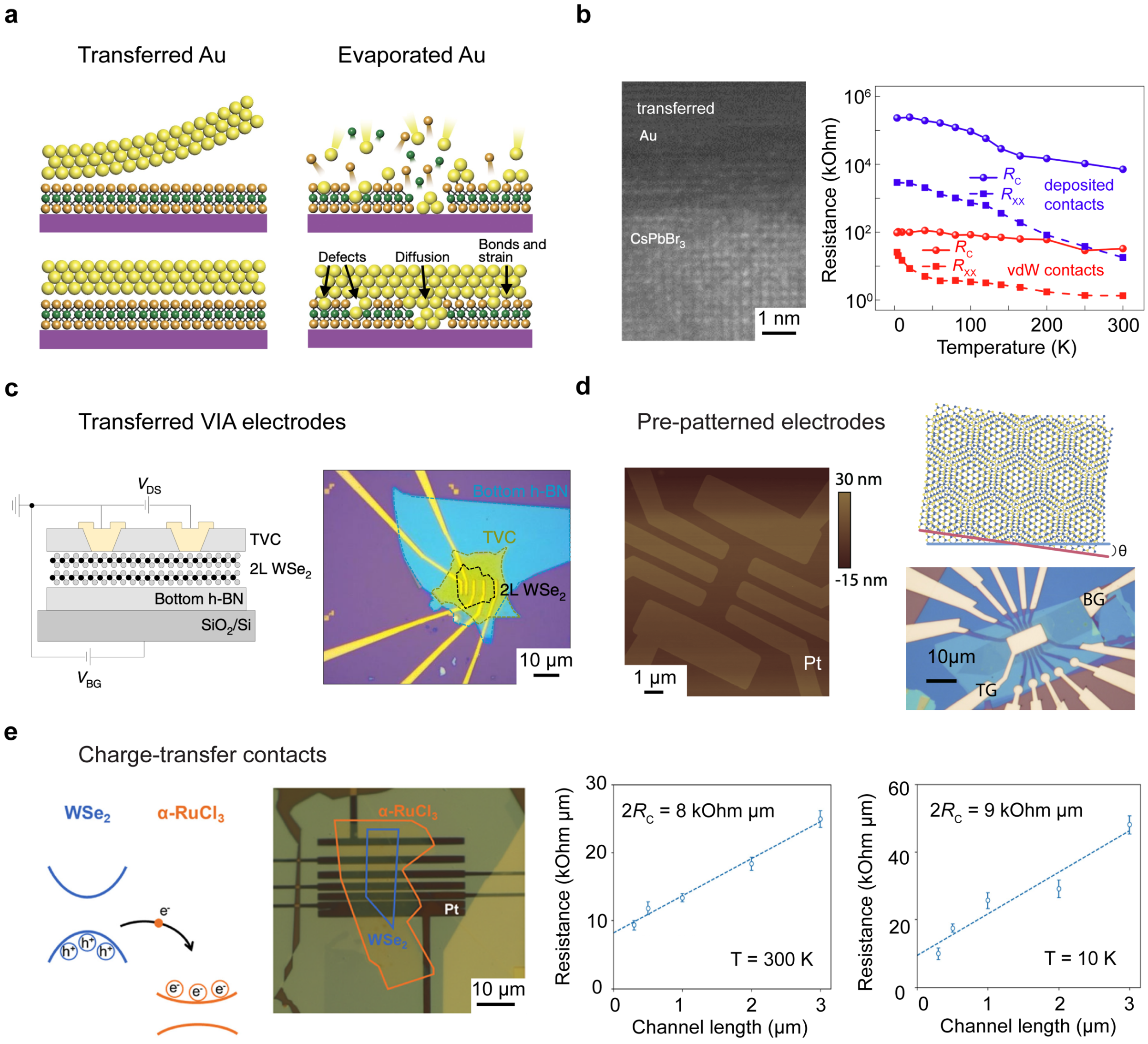

**Figure 7 Transfer-based electrode integration**. (**a**) Cross-sectional schematics of transferred Au electrode (left) and conventionally deposited Au electrodes (right) on top of $MoS_2$. Figures adapted from Ref. [67] Copyright 2018, Springer Nature. (**b**) Scanning transmission electron microscopy image of the cross-section of vdW electrodes / $CsPbBr_3$ interface (left). Comparison of the contact and channel resistance of $CsPbBr_3$ devices with deposited contacts and vdW contacts at various temperatures (right). Figures adapted from Ref. [159] Copyright 2020, Springer Nature. (**c**) Schematic (left) and optical microscopic image (right) of a bilayer $WSe_2$ back-gated transistor with transferred VIA contacts in a TLM arrangement. Figures adapted from Ref. [55] Copyright 2019, Springer Nature. (**d**) AFM image of pre-patterned Pt electrodes on *h*-BN (left). Schematic and optical microscopic image of a twisted bilayer device contacted by pre-patterned Pt (right). Figures adapted from Ref. [80] Copyright 2024, National Academy of Sciences. (**e**) Band alignment of $WSe_2$ and charge-transfer contact α-$RuCl_3$ (left). Optical image of the TLM device (second to the left). Total resistance as a function of channel length from standard TLM measurements at 300 K (second to the right) and 10 K (right), respectively. Figures adapted from Ref. [160] Copyright 2024, American Chemical Society.

The use of transferred electrodes significantly reduced the contact resistance compared to conventionally deposited contacts, thereby allowing for more accurate extraction of intrinsic transport properties (Figure 7b). In these experiments, a hexamethyldisilazane (HMDS) surface treatment procedure was employed to facilitate the delamination of weakly adhesive metals (e.g., Au, Ag, Pt) from Si or $SiO_2$ wafers, resulting in high transfer yield.[67,159] However, this method proves less effective for strongly adhesive metals such as Ti, Ni, and Cr.[67] More recently, a universal method involving a thermally decomposable polymer buffer layer has been developed to further extend the applications of the vdW metal integration approach to strongly adhesive metals or even bulk semiconductors.[161]

In 2019, Jung et al. introduced an interesting contact fabrication strategy utilizing transferred *h*-BN VIA contacts for $WSe_2$ devices, as illustrated in Figure 7c.[55] In this approach, a thin *h*-BN flake (20-30 nm thick) was first patterned and through-etched by RIE to define VIA holes, which were subsequently filled with metal to form contacts. The resulting metallized *h*-BN flake was then used to assemble the 2D device stack, functioning similarly to regular *h*-BN encapsulation while simultaneously contacting the channel from above.[55] Although this approach introduces additional fabrication complexities, it offers a key advantage of precisely defining the contact regions, which is essential for accurate extraction of device transport properties. The low contact resistance and clean channels also enabled the fabrication of high-performance, top-gated $WSe_2$ field-effect transistors, with the *h*-BN flake serving simultaneously as the gate dielectric.[55] Nonetheless, the overall contact quality achieved with the *h*-BN VIA technique is expected to be comparable to that obtained through direct transfer of pre-patterned metal electrodes.

Alternatively, a freshly cleaved 2D flake can be placed directly onto pre-fabricated electrodes on dielectric substrates such as $SiO_2$/Si or *h*-BN,[80] thereby forming electrical contacts from the bottom, as shown in Figure 7d. Similar to the transferred vdW metal strategy, this "back contacting" method also introduces no chemical disorder or physical damage to the TMDs, and it has been widely adopted to fabricate gated transport samples[80,89] and scanning tunneling microscopy devices.[162,163] For example, pre-patterned Pt electrodes (~ 20 nm thick) have been used to successfully contact few-layer twisted $WSe_2$ for transport measurements at cryogenic temperatures,[80] whereas conventional deposition processes often degrade such thin samples and lead to significantly higher contact resistance. It is also important to note that the rigidity of the underlying electrodes necessitates the use of ultra-thin 2D flakes to ensure reliable and conformal contacts, thereby limiting its application to thicker crystals.

More recently, the concept of 'charge-transfer contacts' has been adopted to create low-resistance p-type Ohmic contacts to monolayer $WSe_2$.[160,164] As illustrated in Figure 7e, when $WSe_2$ is brought into contact with a vdW electron acceptor such as $\alpha$-$RuCl_3$, electrons from the valence band maximum of $WSe_2$ spontaneously transfer to the conduction band minimum of $\alpha$-$RuCl_3$, inducing strong hole doping in the contact region. Subsequently, regular electrode materials such as Pt, graphene, or Au can then be applied to the other side of $WSe_2$ to realize effective Ohmic contacts.[160] Using this approach, Xie et al. reported contact resistance as low as 4 kΩ·μm at room temperature and 4.5 kΩ·μm at 10 K for Pt/$WSe_2$ devices, as extracted from TLM measurements. Benefiting from these low-resistance charge-transfer contacts, p-type $WSe_2$ transistors with excellent device performance were demonstrated with ON-current reaching 35 μA/μm and ON/OFF ratios exceeding $10^9$ at room temperature.[160] Furthermore, Pack et al. showed that such contacts could enable electrical transport characterization of monolayer semiconductors at cryogenic temperatures with carrier densities as low as $1.6 \times 10^{11}$ $cm^{-2}$, allowing access to a variety of correlated electronic states.[164]

While various transfer-based contact integration methods have been developed primarily for layered TMDs, certain approaches such as van der Waals metal integration also show strong potential for establishing high-quality contacts to planarized, non-exfoliable crystals. The absence of interface damage in devices using vdW contacts, as evidenced by the significantly improved contact performance of transferred Au on $CsPbBr_3$,[159] underscores the versatility of this method beyond 2D materials. On the other hand, pre-patterned back contacts or charge-transfer contacts are less suitable for bulk crystals with substantial thicknesses, owing to challenges in achieving conformal contact interfaces and uniformly doped regions across the entire crystal, respectively. Continued innovation to create low-resistance contacts to emerging crystals will greatly facilitate their transport studies, further enabling the discovery and manipulation of new quantum phenomena and functionalities in these materials.

## 8. Summary and outlook

In this review, we have highlighted recent advances in the fabrication of multi-terminal electrical contacts on newly developed single crystals for transport measurements. Depending on a material's exfoliability, available sample dimensions, and propensity to degradation, a range of techniques, including mechanical exfoliation, FIB thinning or polymer-based planarization strategies, as well as dielectric encapsulation, shadow mask patterning, and transfer-based contact integration, can be strategically employed, as summarized in Figure 8 and Tables 1 and 2.

These recent technological advances have broadened the range of single-crystalline materials accessible to systematic transport studies, many of which were previously excluded from electrical characterization. Nonetheless, each contact fabrication strategy still presents certain limitations and only works well for specific materials. For example, while the mechanical exfoliation strategy has been widely adopted for layered materials, it often yields small flakes, posing challenges for precise alignment, electrode patterning, and contact optimization. Moreover, for FIB thinning and polymeric planarization methods, despite their ability to accommodate small, non-exfoliable crystals, it is crucial to avoid introducing extrinsic effects such as radiation-induced defects and strain during device fabrication, which may potentially compromise the material's intrinsic properties, particularly for semiconductors and insulators. Air-sensitive crystals further necessitate the use of encapsulation layers or processing in an inert atmosphere for sample handling and transferring before measurements.

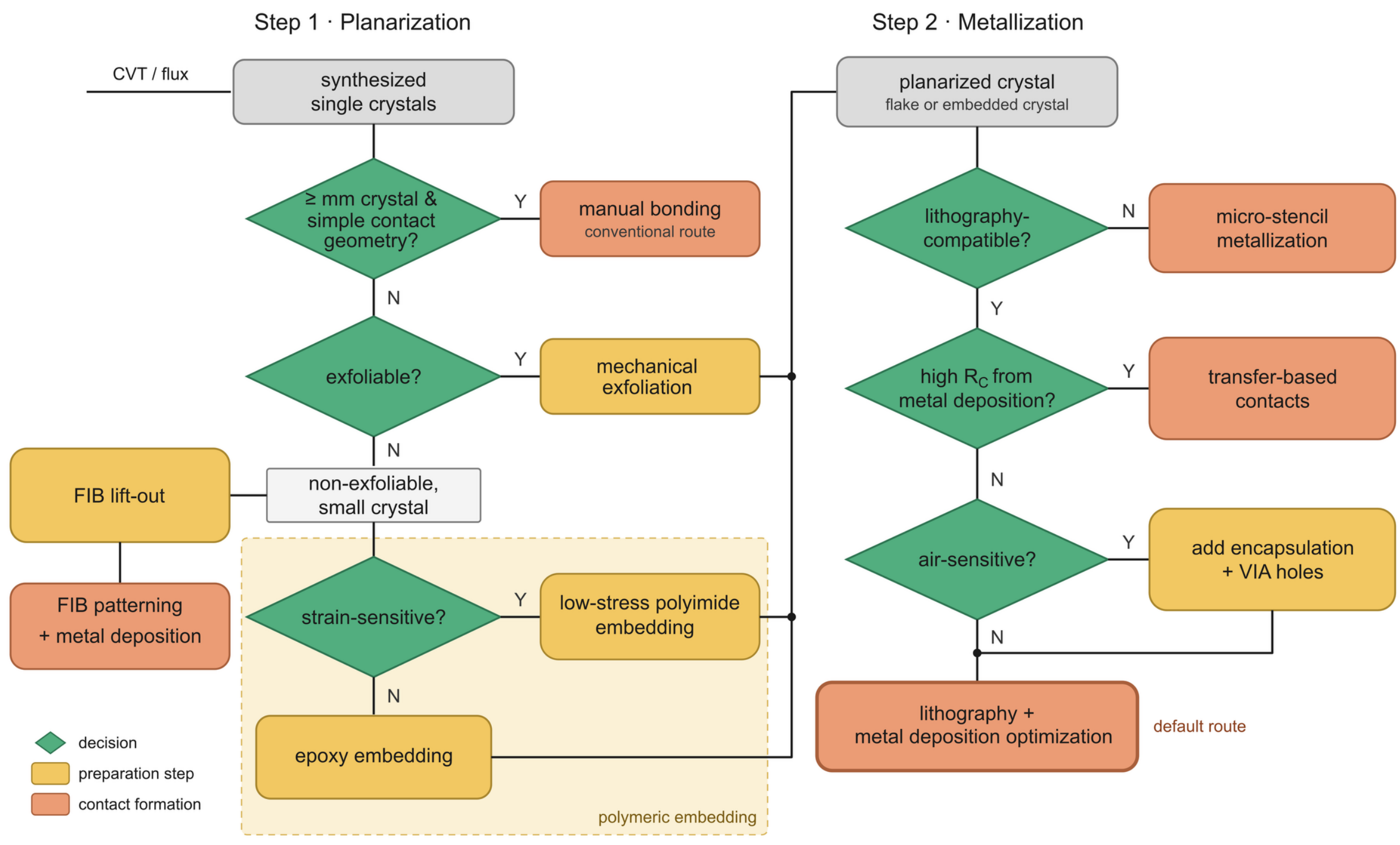

**Figure 8 Decision guide for choosing a contact-fabrication strategy for emerging single crystals**. Green diamonds are decision points, yellow boxes are sample-preparation steps, and orange boxes are contact-formation methods.

Given the diversity of materials and fabrication routes introduced in this review, contact quality is best characterized by several complementary criteria rather than a single one: (1) the contact resistance $R_C$, extracted either from TLM or by comparing two- and four-probe resistances, should remain orders of magnitude smaller than the channel resistance, such that it neither dominates two-terminal measurements nor perturbs four-probe voltage sensing; (2) Ohmic contact behavior should be verified through linear *I-V* characteristics across the measurement conditions of interest, e.g., temperatures and gate voltages; (3) $R_C$ should depend only weakly on temperature[159,160], as a pronounced thermally activated increase upon cooling indicates a Schottky barrier-limited carrier injection; and (4) absolute values of $R_C$, including those cited in Sections 2 and 7, should be compared with caution across different studies, as they can depend strongly on the material and its band alignment with the contact metal, the carrier density, the contact geometry, and the measurement temperature.

**Table 2 Comparison of metallization methods for contact fabrication on emerging single crystals.**

| Metallization methods | Key advantages | Principal limitations | Precautions | Representative examples |
|---|---|---|---|---|
| Manual bonding (conventional route) | No lithography needed; fast; proven in discovering new phenomena | Needs large crystals; ill-defined geometry; no contact optimization | Avoid shorting between wires; poor contacts on gapped systems | Large CVT or flux grown crystals such as $(TaSe_4)_2I$ [3] |
| Lithography + metallization optimization (default route) | Flexible contact geometry (Hall bar, TLM, van der Pauw); $R_C$ can be tuned through optimizing deposition processes | Exposes the crystal to heat, solvents and resist residues; deposition damage and Fermi-level pinning | Verify $R_C$ by TLM before interpreting transport | $MoS_2$ [81], $WS_2$ [53], graphene edge contacts [2] |
| Micro-stencil metallization (alternative) | No high-energy radiation, excessive heat, or chemicals involved; reusable $SiN_x$ and Si micro-stencils; high resolution possible (with Si membrane) | Complex mask fabrication; manual alignment; metal spreading due to poor conformal contact | Add an h-BN cap after deposition if air-sensitive; polymer masks may contaminate the surface | Applicable to most crystals including CeSiI [18] |
| Transfer-based contacts | Damage-free vdW interface with no Fermi-level pinning | Mostly demonstrated on 2D TMDs; limited choice of vdW metals; extra fabrication complexity | Charge-transfer material heavily dopes the contact region | $MoS_2$ [67], $CsPbBr_3$ [159], $WSe_2$ [55] |
| Encapsulation + VIA holes (added before lithography) | Channel stays protected; contact regions are precisely defined | *h*-BN suits flakes only; the polyimide route requires crystal embedding; native oxide must be passivating | Handle in an inert atmosphere until encapsulated; poor RIE selectivity can create edge contacts | 1*T*-$TaS_2$ [130], $BaTiS_3$ [137] |

Beyond contact quality itself, it is important to verify that the device fabrication process does not introduce extrinsic features into the measured transport. Key signatures such as phase transitions, resistance anomalies, or quantum oscillations should persist across multiple devices, different thermal cycles, and, when possible, across devices prepared by different fabrication routes, as exemplified by the progressively shifted or suppressed transitions in epoxy-embedded $BaTiS_3$ crystals discussed in Section 4.[6] Comparison with as-grown bulk crystals, using conventional manually bonded reference samples, thermodynamic probes such as heat capacity, and complementary structural and surface characterization, can further distinguish intrinsic behavior from fabrication-induced strain or damage.

A broader limitation lies in the imbalanced technological maturity across different classes of emerging crystals. To date, most existing contact-optimization efforts and systematic transport studies have focused on layered materials, whose device fabrication protocols are relatively well established. In contrast, for many newly synthesized, non-exfoliable crystals, research efforts have primarily focused on establishing working electrodes, either through the conventional manual bonding method or lithography-based microfabrication processes building on an appropriate planarization strategy, to enable preliminary transport measurements. Continued research efforts are therefore needed to further optimize contact fabrication procedures for non-exfoliable crystals, in a manner comparable to conventional semiconductors and 2D materials. Such advances are essential for unlocking the full potential of these emerging materials and will be pivotal for deepening our fundamental understanding of intriguing physical phenomena and enabling next-generation device technologies.

## Biographies

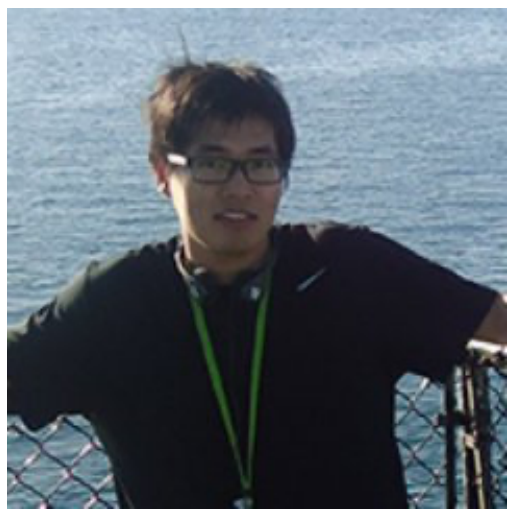

**Huandong Chen** is currently a postdoctoral researcher in the Condensed Matter Physics and Materials Science Department at Brookhaven National Laboratory, working on low-dimensional quantum materials and devices using STM. He received his Ph.D. degree in Materials Science from University of Southern California in 2023, where he has worked on bulk single crystal synthesis, transport studies and electronic and optoelectronic devices of novel complex chalcogenides.

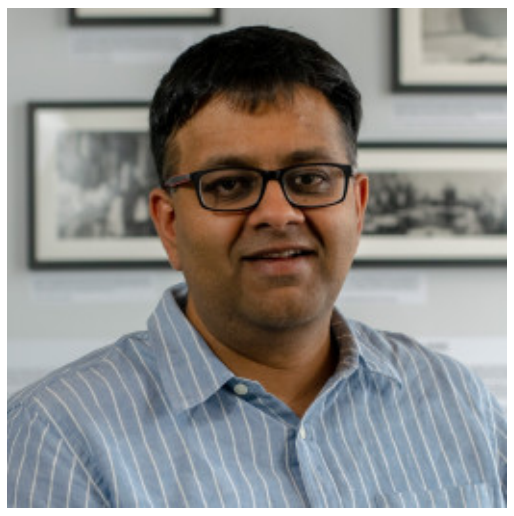

Abhay Narayan Pasupathy is currently a Professor of Physics at Columbia University and a Group Leader in the Condensed Matter Physics and Materials Science Division at Brookhaven National Laboratory. His group works on experiments to understand the quantum properties of emerging materials using scanning probe microscopy techniques and electron transport measurements.

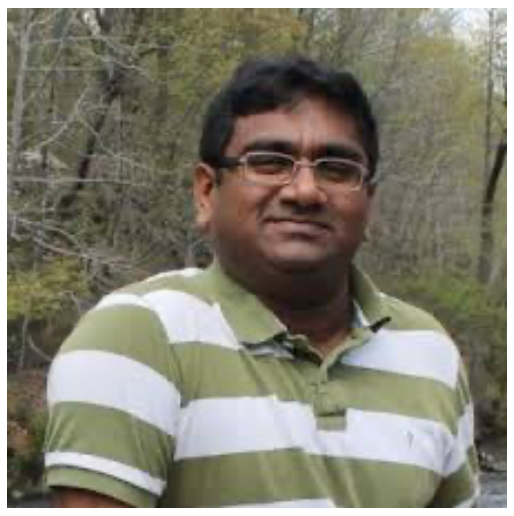

**Jayakanth Ravichandran** is currently a Professor of Chemical Engineering and Materials Science and Electrical and Computer Engineering at University of Southern California. He is also a co-director of the Core Center of Excellence in Nano Imaging and an associate editor for Journal of Materials Research. He completed his graduate work at University of California, Berkeley in Applied Science and Technology, and postdoctoral work in Physics at Columbia University and Harvard University. His research interests are in the broad area of electronic and photonic materials and devices.

# Author declarations

## Author contributions

J.R. supervised the project. H.C. carried out literature survey on the topic. H.C. wrote the manuscript with input from all other coauthors.

## Funding

We gratefully acknowledge support from an ARO MURI program (W911NF-21-1-0327), an ARO grant (W911NF-24-1-0164) and National Science Foundation (DMR-2122071). The work at Brookhaven National Laboratory (BNL) is supported by the Office of Basic Energy Sciences, Materials Sciences and Engineering Division, U.S. Department of Energy (DOE) under contract No. DE-SC0012704.

## Conflict of interest

The authors declare no competing financial interests.

## Data availability statement

Data sharing is not applicable to this article.